\documentclass[epj]{svjour}

\usepackage{graphicx}
\usepackage{fancyhdr}
\def\makeheadbox{{
 }}
\def\mytitle{My title} 
\def\myauthors{My name}  
\def\mytype{My type of session}
\def\mysession{My session}

\def\mytitle{Search for new physics in top events with the D{\O} detector} 
\def\myauthors{Isabelle Ripp-Baudpt}    
\def\mytype{Contributed Talk}    
\def\mysession{Alternatives}

\begin{document}
\title{Search for new physics in top events with the D{\O} detector}
\author{Isabelle Ripp-Baudot\inst{1}
\thanks{on behalf of the D{\O} collaboration}
}                     
%
%
\institute{IPHC, Universit\'e Louis Pasteur, CNRS/IN2P3, Strasbourg, France}
%
\date{}
%
\abstract{
This review is focused on the search for new processes, performed with top quark events in D{\O}.
It presents four updated or new D{\O} results. The two first analyses deal with top production 
properties: they search for a new heavy resonance decaying to $t \bar t$.
The two last results concern top decay properties:
the measurement of the $W$ helicity as a probe of the $tWb$ coupling structure, and the 
top quark branching ratio to $Wb$. Neither of these measurements reveal any deviation with respect 
to the standard model predictions. 
}

\maketitle
%
\section{Introduction}

\label{intro}

This review is focused on the search for new physics performed with top quark events in D{\O}.
It presents four updated or new D{\O} results. The two first analyses deal with top production 
properties: they search for a new heavy resonance decaying to $t \bar t$.
The two last results concern top decay properties:
the measurement of the $W$ helicity as a probe of the $tWb$ coupling structure, and the 
top quark branching ratio to $Wb$.

\bigskip

The Tevatron $p \bar p$ collider at Fermilab operates with a center-of-mass energy of 1.96 TeV. Beams are
colliding in two points instrumented by the CDF and the D{\O} multipurpose detectors. Since 2001 the Tevatron is
in its second run and, thanks to very good performances, it has already delivered more than 3 $\mbox{fb}^{-1}$
to each experiment in 2007. The results included in this review are those based 
on about 1 $\mbox{fb}^{-1}$ of data. Up to 
4 to 8 $\mbox{fb}^{-1}$ are expected per experiment until the 
smooth running of the LHC, i.e. at least 50 times as much
top as available at the end of the Run I which lead to the top discovery.

Up to now the top quark is observed experimentally via $t \bar t$ pair production through the strong interaction.
Electroweak production of a single top is also possible but with a smaller cross-section and a larger background,
so that the first evidence of this process has been shown only recently by D{\O} \cite{singletop}.
Single top events are also sensitive to new physics but this channel suffers
from its low statistics. All analyses presented in the
following are therefore performed with $t \bar t$ events.

Due to its high mass, the top quark decays to a real $W$ and a b quark before hadronizing.
Measurements are grouped according to the $W$ decay channel. The more leptons present in the final state, the
smaller the background but also the lower the statistics. Therefore the considered channel will generally be
 the lepton+jets one, accounting for about 30 \% of the $t \bar t$ final states. The background for this
 channel can be split into two components: on the one hand a physical background originating from processes
 showing a similar final state, among which the dominant one is the $W$+jets production; on the other hand 
 an instrumental background coming from the multijet production where a jet mimics the isolated lepton.
 The background is efficiently reduced  by requiring one or two jets to be b-tagged, using track impact
 parameters and displaced vertices information combined in a neural network algorithm \cite{nnbtag}.

\section{Top production properties}
\label{sec:1}

\subsection{New resonance direct search}
\label{sec:11} 

D{\O} has performed a model independant search 
for a new heavy neutral resonance decaying to $t \bar t$ \cite{ttresonance}, generically
referred to as $Z'$. The $Z' \to t \bar t$ channel is of particular interest in models with a leptophobic
$Z'$ decaying dominantly to quarks. The width of this $Z'$ boson is assumed
to be narrow, that is to say small with respect to the detector resolution, so that the
resonance should be visible in the $t \bar t$ invariant mass distribution.

Several methods have
been evaluated to reconstruct this invariant mass. The chosen one combines the cha\-r\-ged 
lepton momentum, the four
leading jets momenta and the inferred $\nu$ momentum.
The $t \bar t$ invariant mass distribution is used to
perform a binned likelihood fit of the signal and background expectations compared to data. 

\begin{figure}[h!]
\includegraphics[width=0.50\textwidth,height=0.30\textwidth,angle=0]{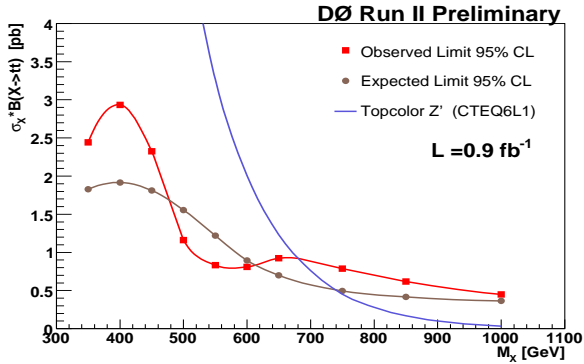}
\caption{Expected and observed upper limits on the heavy resonance production cross-section time 
the branching ratio to $t \bar t$ as a function of the resonance mass.}
\label{fig:ttresonance}      
\end{figure}

No evidence for a  $t \bar t$ narrow resonance has been found.
Upper limits on the heavy resonance production cross-section time the branching ratio to $t \bar t$ are
extracted over all the range of resonance masses. They are shown in figure \ref{fig:ttresonance}.
Within a topcolor assisted technicolor model, the existence of a narrow leptophobic $Z'$ decaying to 
$t \bar t$ is ruled out at the 95 \% C.L. below 680 GeV/$\mbox{c}^2$.

\subsection{Top forward-backward asymmetry} 
\label{sec:12}

The measurement of the forward-backward asymmetry in the $t \bar t$ pair production at Tevatron is 
sensitive to new physics processes, as for example an alternative pair production via the decay 
of a new heavy neutral boson.
This study does not need to assume anything about the resonance width, and thus it is complementary 
to the direct search of a narrow resonance presented in section \ref{sec:11}. It is sensitive to 
any new process giving a $t \bar t$ final state with a large forward-backward aymmetry, 
as the standard model predicts only a small
asymmetry in the $t \bar t$ pair production at Tevatron

This forward-backward asymmetry is measured in D{\O}
from the signed difference between the $t$ and the  $\bar t$ rapidities. 
The acceptance may strongly shape the observed asymmetry, mostly through criteria imposed on the
number of jets and on the jet momenta. The asymmetry is furthermore diluted by 
reconstruction effects such as misidentification of the lepton charge. 
In order to facilitate comparisons with theoretical calculations, 
the analysis is designed to have an
acceptance which can be easily described at the particle level. Thus the obtained result is given 
for a given region of phase space, specific to this analysis, and uncorrected for
reconstruction effects. 

\begin{figure}[h!]
\includegraphics[width=0.50\textwidth,height=0.30\textwidth,angle=0]{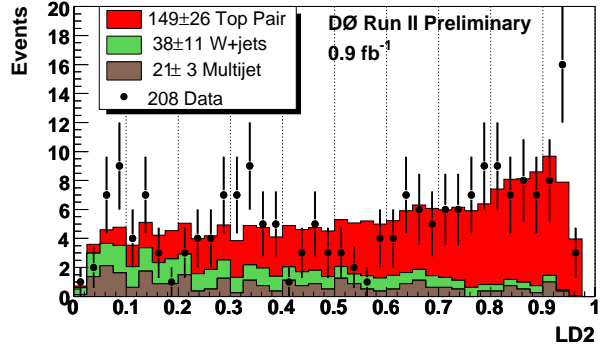}
\caption{Distribution of the multivariate discriminant function for the data (dots)
and the model (histogram) for forward top quark events.}
\label{fig:asym1} 
\end{figure}

\begin{figure}[h!]
\includegraphics[width=0.50\textwidth,height=0.30\textwidth,angle=0]{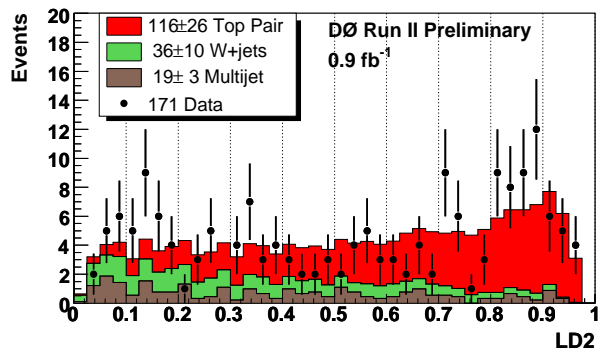}
\caption{Distribution of the multivariate discriminant function for the data (dots) and the model (histogram)
for backward top quark events.}
\label{fig:asym2} 
\end{figure}

The measured asymmetry is \cite{ttasymmetry}

\noindent
$A_{FB} = 12 \pm 8 (\mbox{stat.}) ^{+1.1}_{-1.0} (\mbox{syst.})$ \%, in good agreement
with the standard model predictions. This measurement is obtained through a likelihood fit comparing the
distribution of a multivariate function to templates for forward and backward top quark events. 
These distributions are shown in figures \ref{fig:asym1} and \ref{fig:asym2}.
The extracted upper limits on $A_{FB}$ are translated into limits on the
fraction of top-pair events produced via the decay of a heavy boson. The high measured
asymmetry value does not yield
yet useful limits on this production channel and it will be interesting to do this analysis with more statistics.

\section{Top decay properties}
\label{sec:2}

\subsection{Top-W-b coupling structure}
\label{sec:21}

In the standard model the charged weak current is described by a pure V-A structure. This fundamental property
can be tested at the highest energy scale by measuring the $W$ boson helicity in the top decay. The standard
model predictions for the three helicity fractions, at the tree level and in the massless b quark approximation,
are 30 \% of left-handed $W$, 70 \% of longitudinal $W$ and no right-handed $W$ 
(i.e. expected right-handed fraction $f_+$ = 0). These fractions are measured in D{\O} from the
angular distribution of the top decay products. The considered angle is between the charged lepton direction in
the $W$ restframe and the $W$ direction in the top restframe.

\begin{figure}[h!]
\includegraphics[width=0.50\textwidth,height=0.30\textwidth,angle=0]{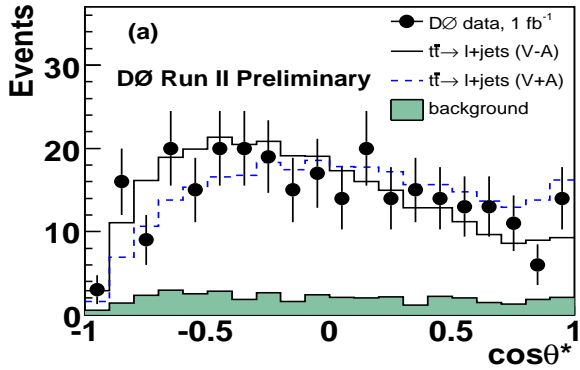}
\caption{Angular distribution for the lepton+jets data and for the predicted sum of signal and 
background in the standard model and assuming a pure V+A structure.}
\label{fig:whel_ljets} 
\end{figure}

\begin{figure}[h!]
\includegraphics[width=0.50\textwidth,height=0.30\textwidth,angle=0]{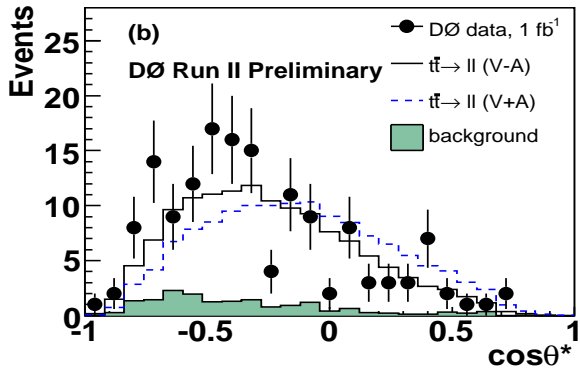}
\caption{Angular distribution for the dilepton data and for the predicted sum of signal and 
background in the standard model and assuming a pure V+A structure.}
\label{fig:whel_dil} 
\end{figure}

For this measurement only, both the lepton+jets and the dilepton channels are used. Events are selected in two
steps. After a set of sequential cuts, 
a multivariate discriminant function is formed using both kinematical and b-tagging
informations. The composition in signal and background of the selected candidates is obtained by
fitting the distribution of this discriminant function. Then a cut on this multivariate
discriminant variable further enriches the sample in 
$t \bar t$ signal. The reconstructed angle is estimated through a constrained fit. A binned likelihood compares
this angle reconstructed in data
to templates formed with different fractions of a non-standard V+A admixture to the weak current.
The fraction of longitudinal $W$ is fixed to its standard value of 0.7 during the fit. The obtained result 
for the fraction of right-handed $W$ is \cite{whelicity}
$f_+ = 0.017 \pm 0.048 (\mbox{stat.}) \pm 0.047 (\mbox{syst.})$,
 in good agreement
with the standard model prediction. Figures \ref{fig:whel_ljets} and \ref{fig:whel_dil} show the angular
distribution for the data as well as 
for two simulations: the predicted sum of signal and background in the standard model and assuming a
pure V+A structure. 

The future plan for this analyses
is to fit simultaneously both fractions of longitudinal and
right-handed $W$, giving a model independant result.

\subsection{Top quark branching ratio}
\label{sec:22}

This measurement tests the branching ratio of the top quark to the b quark with respect to all other known
down-type quarks. In the standard  model we define the ratio $R = {\cal B}(t \to Wb) / {\cal B}(t \to Wq)$, 
where $q$ refers to
any known down-type quark. This ratio $R$ can be expressed in terms of the corresponding CKM
matrix elements. Once it is precisely measured it will provide a model independant measurement of the
matrix element $|V_{tb}|$. 
A deviation of $R$ from unity would indicate non-standard top decay. But to be sensitive to such
effects at the Tevatron requires very good accuracy for this measurement, at the \% level, 
and may need the combination of CDF and D{\O} results.

Experimentally $R$ can be determined simultaneously with the pair-production cross-section 
$\sigma_{t \bar t}$ by classifying the 
$t \bar t$ selected candidates into events with 0, 1 and 2 b-tagged jets. A binned likelihood fit is performed 
which compares the observed number of events in these 3 bins with the prediction for signal and background.
A discriminant function based on the kinematic properties of the $t \bar t$ events is used to further constrain
the number of events without b-tagged jets.
The dependance on $R$ of the selection and the b-tagging efficiencies is taken into account when
predicting the signal contribution in each bin. 
The values of the the ratio $R$ and of the cross-section that maximize the
2-dimensional likelihood function are \cite{tbratio} 

\noindent
$R= 0.991 ^{+0.094}_{-0.085} (\mbox{stat.+syst.})$ and 

\begin{figure}[h!]
\includegraphics[width=0.48\textwidth,height=0.35\textwidth,angle=0]{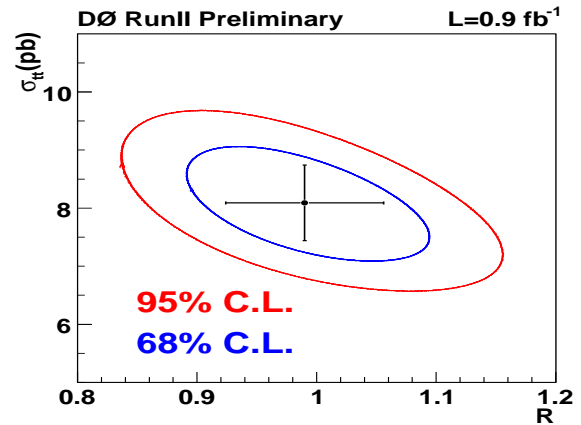}
\caption{Values of the simultaneously measured ratio $R$ and cross-section $\sigma_{t \bar t}$ 
with the 1 and 2$\sigma$ contours.}
\label{fig:ratioBR}
\end{figure}

\noindent
$\sigma_{t \bar t} = 8.10 ^{+0.87}_{-0.82} (\mbox{stat.+syst.}) \pm 0.49 (\mbox{lumi.})$ pb, 
they are illustrated in figure 
\ref{fig:ratioBR}. These values are in good agreement with the standard model expectations of 
$R = 1$ and $\sigma_{t \bar t} = 6.8 \pm 0.6$ pb \cite{xsection} for a top quark mass of 175 $\mbox{GeV/c}^2$.
This measurement of $R$ is by far the most precise up to now.

\section{Conclusion}
\label{conclu}

The top quark is the heaviest fundamental known particle and its study allows us to test the standard
model at the highest energy scale available. 
The Tevatron has been up to now the unique place to study the top quark.
With the Run II data we have entered a phase of precision measurements in the top sector.
This precise characterization of the top quark aims at discovering hints of the physics laying
beyond the standard model. In this review we have reported four analysis performed by D{\O} and
looking for new processes occuring during the top production and the top decay.
Neither of these measurements reveal any deviation with respect to the standard model predictions. 
Nethertheless they are still statistically limited and they will benefit from the expected increase
of the luminosity in the next two years.


\end{document}